\documentclass[12pt]{article}
\usepackage{epstopdf}
\usepackage{subfigure}
\usepackage{lscape}
\usepackage{natbib}
\usepackage{amssymb}
\usepackage{amsthm}
\usepackage{amsmath}
\usepackage{xcolor}
\usepackage[colorlinks, linkcolor=red, citecolor=blue, urlcolor=blue]{hyperref} 
\usepackage{eurosym}
\usepackage{amsfonts}
\usepackage{booktabs,caption}
\usepackage{graphicx}
\usepackage{bbm}
\usepackage{array}
\usepackage{calc}
\usepackage{tabularx}
\usepackage{multirow}
\usepackage{arydshln}
\usepackage{setspace} 
\edef\restoreparindent{\parindent=\the\parindent\relax}
\usepackage{parskip}
\usepackage[titletoc,title]{appendix}
\restoreparindent

 \usepackage[UKenglish]{babel}
 \usepackage[UKenglish]{isodate}

\definecolor{dgreen}{rgb}{0.,0.6,0.}

\newcolumntype{L}[1]{>{\raggedright\arraybackslash}p{#1}}
\newcolumntype{C}[1]{>{\centering\arraybackslash}p{#1}}
\newcolumntype{R}[1]{>{\raggedleft\arraybackslash}p{#1}}
\providecommand{\U}[1]{\protect\rule{.1in}{.1in}}

\newcommand{\killpunct}[1]{}

\usepackage[a4paper,top=20mm, bottom=24mm, width=168mm]{geometry}

\usepackage{titlesec}
\titleformat{\section}{\Large\bfseries}{\thesection . }{0.5em}{}
\titleformat{\subsection}{\large\bfseries}{\thesubsection }{0.5em}{}

\titlespacing{\section}{0pt}{\parskip}{\parskip}

\emergencystretch=\maxdimen
\usepackage{placeins}

\renewcommand\footnotemark{}
\begin{document}

\sloppy

\title{Gambling on Momentum}

\author{Marius \"{O}tting \\ \small{Bielefeld University} \and Christian Deutscher \\ \small{Bielefeld University} \and Carl Singleton \\ \small{University of Reading} \and Luca De Angelis\thanks{\"{O}tting: Deparment of Business Administration and Economics and Department of Sport Science, Bielefeld University, Germany (email: \href{mailto:marius.oetting@uni-bielefeld.de}{marius.oetting@uni-bielefeld.de}); corresponding author. Deutscher: Department of Sport Science, Bielefeld University, Germany (email: \href{mailto:christian.deutscher@uni-bielefeld.de}{christian.deutscher@uni-bielefeld.de}).\ De Angelis: Department of Economics, University of Bologna, Italy (email: \href{mailto:l.deangelis@unibo.it}{l.deangelis@unibo.it}).
Singleton: Department of Economics, University of Reading, Whiteknights, RG6 6EL, UK (email: \href{mailto:c.a.singleton@reading.ac.uk}{c.a.singleton@reading.ac.uk}).
\newline \indent 
\newline \indent Declarations of interest: none}\\ \small{University of Bologna} 
\\
}
\date{November 2022}

\maketitle

\begin{abstract}
\noindent Sports betting markets are proven real-world laboratories to test theories of asset pricing anomalies and risky behaviour. Using a high-frequency dataset provided directly by a major bookmaker, containing the odds and amounts staked throughout German Bundesliga football matches, we test for evidence of momentum in the betting and pricing behaviour after equalising goals. We find that bettors see value in teams that have the apparent momentum, staking about 40\% more on them than teams that just conceded an equaliser. Still, there is no evidence that such perceived momentum matters on average for match outcomes or is associated with the bookmaker offering favourable odds. We also confirm that betting on the apparent momentum would lead to substantial losses for bettors.
\end{abstract}

\noindent \textit{Keywords}: Behavioural bias, Betting markets, Market efficiency, Momentum, Risk-taking\\
\newline
\noindent \textit{JEL codes}: G14, G41, L83, Z2

\clearpage
\doublespacing
\renewcommand\footnotemark{1}
\sloppy
\section{Introduction}
\noindent Human performance is rarely perfectly consistent over time. Instead, the same workers and their teams can give outstanding performances on some days but be rubbish on others. In striving to predict the future, humans often fixate on salient patterns in historical performance data, such as streaks of success or failure, even if the underlying data-generating process is somewhat driven by white noise and randomness. One such fixation is the concept of the ``hot hand'', according to which there is a belief that serial correlation exists in human performance. This concept has been analysed often in sports. Since the seminal study by \citet{gilovich1985hot}, who did not find support for the hot hand in professional basketball shooting, the evidence on whether such beliefs could be correct has been generally mixed (e.g., \citealp{green2017hot,miller2016surprised,otting2020hot,tversky1989cold,wetzels2016bayesian}).

While many studies investigate a potential hot hand effect in sports, this notion has also been applied to other settings, especially finance; strategies that attempt to identify, follow and profit from market trends are popular among professional investors. In the finance literature, there is abundant evidence that this ``momentum'' explains some of the cross-section of returns in financial markets
\citep{Asness2013,Barroso2015,Carhart1997,ehsani2022factor,Fama2012,Jegadeesh2011,Novy2012}, 
as well as defining profitable investment strategies in time series frameworks \citep{chan1996momentum,Moskowitz2012,Zakamulin2022}.
However, as typically occurs in finance due to the complexity of markets and the multitude of factors that drive asset prices, there are also some findings in the literature that do not support the existence of momentum \citep{Huang2020,Kim2016}.

When testing behavioural asset pricing theories, sports betting markets constitute a useful real-world laboratory with many advantages over traditional financial markets \citep{bar2020ask,Thaler1988}. The terminal value of sports betting contracts is exogenous to investor behaviour and provides clean identification of mispricing, unlike in traditional financial markets. In particular, if prices deviate from fundamentals due to cognitive biases or erroneous beliefs among the market participants, they will be corrected on average by the sporting outcomes, which are exogenous to these biases or beliefs \citep{MOSKOWITZ2021}. Biases investigated in betting markets include the overvaluation of longshots (e.g., \citealp{Angelini2019,OTTAVIANI2008favlong,Vlastakis2009}) and overreactions to major in-play events such as goals (e.g., \citealp{Angelini2022,Choi2014,Croxson2014}).

In this paper, we have unusual access to high-resolution betting market data provided by a large European bookmaker. We use these data to investigate how bettors respond to perceived momentum. The data cover second-by-second betting odds and volumes staked on all potential outcomes for two seasons of the German Bundesliga, covering 612 football matches, enabling a clean analysis of the responses to momentum within betting markets. We focus on the equalising goal in matches that have a 1-1 scoreline since this event implies some notion of apparent momentum. Although equalisers reset the two teams to their relative position at kick-off, the team that recently scored the equaliser might be considered more likely to score another goal due to the momentum gained by scoring the equaliser. For these 1-1 scorelines, we investigate the betting activity in the minute after the equalisers. We find no evidence that the sequence of goals on average impacts the winning chances of teams or the betting odds provided by the bookmaker. However, we find that bettors have a strong tendency toward betting on the team that has gained momentum. The traded volumes placed are considerably higher on teams that scored the equaliser, compared to teams that conceded at 1-1. In particular, stakes placed on teams that scored the equaliser are about 40\% higher compared to teams that just conceded the equaliser. We further find that a corresponding betting strategy of always following this momentum, or believing it could be profitable, would in fact yield substantially negative returns to bettors.

Correspondingly, in financial markets, trade volumes have been shown to have a fundamental link with price momentum \citep{lee2000price}. Closely related to our study, \cite{Levitt2004}, in his seminal study on the economics of gambling markets, analysed some data akin to real-world betting stakes, but these related to a pre-match season-long prediction competition for American football, with an entry fee and no cost per bet, for a relatively small number of selected participants, and thus did not cover general market activity. \citeauthor{Levitt2004} though used these data to challenge the balanced book hypothesis, whereby bookmakers would just adjust their prices according to the flow of bets and eliminate risk from their position in the market. \citeauthor{Levitt2004}'s results instead suggested that bookmakers may not only be good at forecasting outcomes but also know the likely behavioural biases of their customers, which they can exploit to maximise profits. 

Our paper contributes to the literature on behavioural asset pricing in financial markets, by investigating behavioural biases in high-frequency (in-play) betting markets. 
There are other contributions in the literature that analyse some sort of momentum effects in betting markets (e.g., \citealp{brown1993does,camerer1989does,Durand2021,Krieger2021,Metz2022,paul2005bettor,wheatcroft2020profiting,woodland2000testing}), but at a much lower frequency level of data, through the long-term patterns in final event outcomes, typically covering weeks or months, and mostly focusing only on posted prices (odds) and not volumes (stakes). \cite{Paul2014hothand}, however, provide the exception, being to the best of our knowledge the only study to use actual data on betting action from bookmakers to study the ``hot hand'' in sports. They found that the volume of pre-game betting activity on National Football League matches followed teams on a hot streak of wins and avoided those on a losing streak. Our study is the first to consider momentum and betting activity at a much higher frequency and immediacy.

The rest of the paper is structured as follows. Section 2 introduces the unique high-frequency betting market dataset. Section 3 empirically
investigates the impact of momentum on match outcomes and betting markets. Section 4 concludes. Several robustness checks are provided in the Appendix.

\section{Data}
\noindent Our dataset was provided by a major European bookmaker that has a large customer base in Germany. It covers second-by-second betting odds and volumes for all 612 German Bundesliga football matches in the 2017/18 and 2018/19 seasons, including in-play information about the timing of major events (such as goals and red cards). The betting stakes (amounts) in the dataset have been multiplied by the same constant for all matches since we do not have the bookmaker's permission to represent the true amount of monetary units. Regardless, our data allow us to compare the volumes of betting stakes across and within matches without providing the actual values or statistics of true stakes.

To investigate how bettors (the bookmaker) respond to teams seemingly having momentum, we investigate their betting behaviour (odds-setting process) after the equalising goal for the scoreline 1-1.\footnote{Football is a low-scoring game, and in the entire history of professional football, 1-1 has been the most likely final outcome of a match \cite{reade2021evaluating}.} Equalisers reset the teams to the position they were in prior to the match beginning. Nevertheless, it is possible that the identity of whichever team scored the most recent goal may affect the subsequent behaviour of all the agents involved. In other words, the sequence of goals up to and including the equaliser may help to predict what will happen in the remainder of the match, conditional on pre-match expectations. Both of these possibilities would imply some notion of momentum in a football match, whereby scoring a goal suggests that a team is more likely to score further goals than beforehand. 

As betting volumes vary across teams, we model the relative stakes placed at a moment or over some period of time, where we consider the proportion of the total betting volumes that were placed on a particular team to win the match. In particular, we focus on the stakes placed, winning chances, and betting prices during the next minute after a 1-1 equaliser was scored. For the 612 matches considered, 233 had an intermediate scoreline of 1-1. If a 1-1 equaliser is scored fairly late in a match, such as in the 85th minute or later, much lower absolute stakes tend to be placed in the market due to there being little time left to play, and thus relative stakes across the three match outcomes also become noisy. If the 1-1 is scored during injury time, after the regular 90 minutes of play are complete, then the market does not necessarily reopen again at all (depending on the amount of injury time that the referee has awarded). We thus consider only observations where the equaliser is scored before the 85th minute, resulting in 212 match observations in our sample.

\begin{figure}[ht!]
\centering
\caption{Schalke 04 vs VfL Wolfsburg, 20 January, 2019: Example time series of in-play relative stakes placed on a win for the home team (Schalke), a draw, and the away team (Wolfsburg).}
\includegraphics[scale=1]{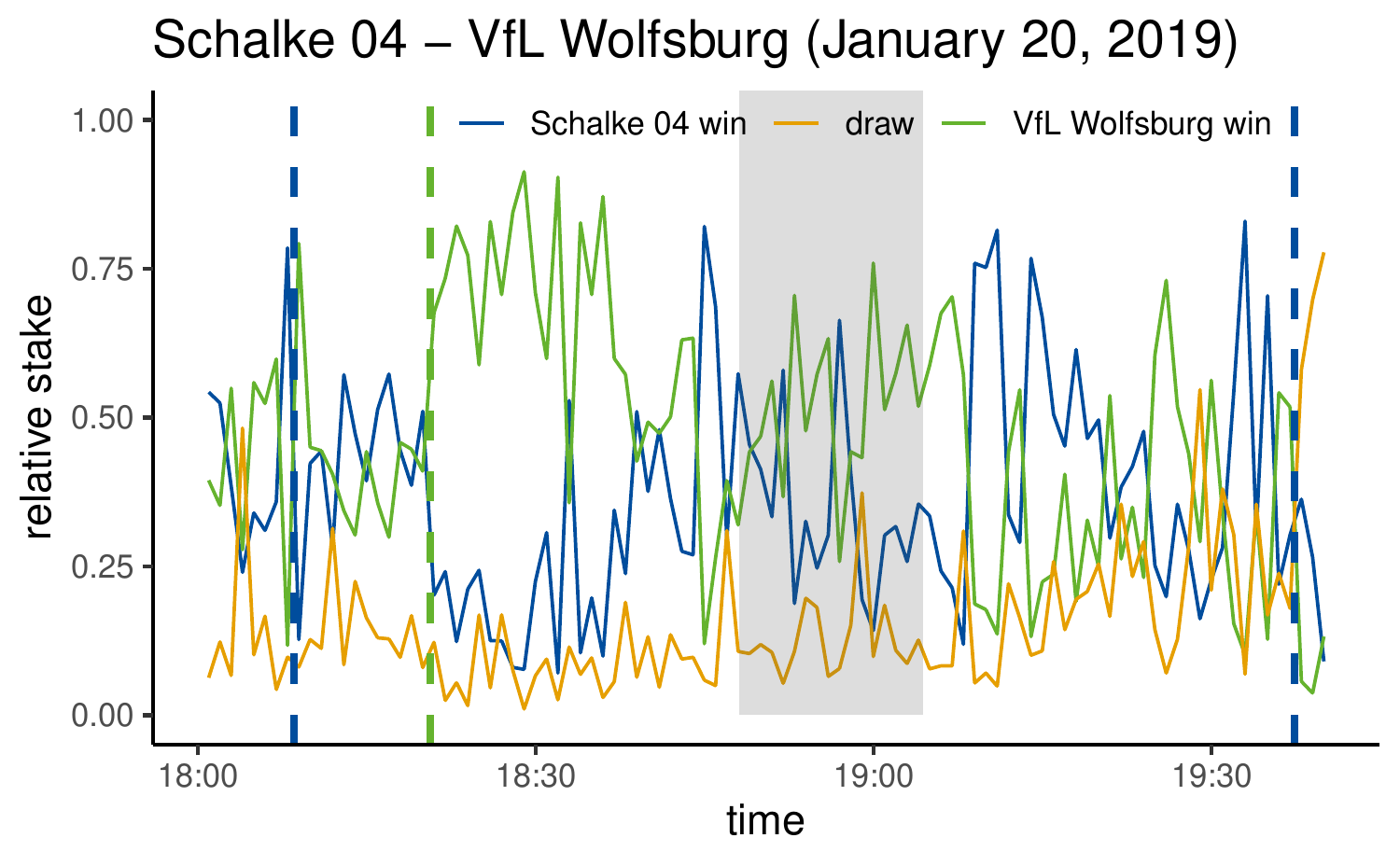}
\caption*{\footnotesize Notes.- The vertical dashed lines denote when goals were scored by Schalke 04 (blue lines) and VfL Wolfsburg (green lines). The grey shaded area indicates half time.}
\label{fig:example_data}
\end{figure}

Figure~\ref{fig:example_data} shows an example minute-by-minute time series of relative stakes from our data for a match between Schalke 04 and VfL Wolfsburg, which kicked off at 18:00 CEST on January 20, 2019, and ended 2-1. Before the match began, the bookmaker prices suggested it would be closely fought, with decimal odds of 2.25 for the home team and 3.0 for both the away team and the draw outcome. This example also gives a first glance at how bettors respond to goals and whatever pricing strategy the bookmaker follows in their aftermath. Slightly higher relative stakes were placed on the less-favoured away team, Wolfsburg, than on Schalke early in the match, while the stakes placed on Schalke increased after they scored the first goal of the match. When Wolfsburg scored the equaliser, the relative stakes placed on Wolfsburg to win increased, and are even larger compared to the early stages of the match before the first goal. As introduced above, such betting behaviour may well be driven by bettors believing there is value in momentum for the equalising team. For all matches that featured a 1-1 scoreline, we observe in the minute after, on average, 47\% of the stakes placed on the equalising team, 37.2\% placed on the conceding team, and only 18.6\% on the draw. However, for these matches, the draw is the most likely final outcome (39.3\%). A defeat for the equalising team was the second most likely outcome (33.2\%), followed by a win for the equalising team (27.5\%). These descriptives give a first impression that bettors believe there is value in the apparent momentum, although the win for the equalising team is the least likely outcome.

In the example match, the absolute betting activity increased after the equaliser. For the time from the kick-off, until the first goal was scored, there was an average amount staked of 47 per minute (transformed values). During the three minutes before the 1-1 equaliser, the betting activity was slightly reduced, with an average amount staked of 30. The amount staked climaxed in the minute after the equaliser at an average of 60 per minute. However, the betting activity quickly reduced thereafter, with an average stake per minute of 49 in the 10 minutes following the equaliser.

The main variables of interest in our analysis, which are introduced in the following, are summarised in Table~\ref{tab:descriptives}. As bettors are more likely to wager their money on favourites, we consider a team's odds-implied probability (\textit{impprob}) of winning at the kick-off, which is derived by taking the inverse of the posted decimal odds for them to win, and normalising this such that it sums to one with the inverse of the odds for the other two possible match outcomes. Teams conceding the equaliser generally have a higher implied probability at the kick-off, unsurprisingly indicating that favourites are more likely to first have a lead and then concede an equalising goal in matches that get to 1-1. This is also in line with the finding from above that a defeat is more likely than a win for an equalising team. If an equaliser is scored a few minutes before the final whistle, then bettors may be unlikely to place their money on a team to win but rather on a draw. We thus consider the \textit{minute} of the equaliser, which is, on average, minute 47, but also occurs in our sample as early as the 5th minute and as late as the 84th minute. As undermanned teams have a reduced chance to win a match, we consider red cards received before the 1-1 equaliser.\footnote{A red card, or sending off, is a relatively rare but severe punishment in a football match, with the receiving team having to play with one fewer player for the remainder of the match.} In particular, the explanatory variable \textit{redcard} gives the difference in the number of red cards received between the equalising and conceding teams before a match arrives at 1-1, which in our sample lies strictly between minus and positive one.
\begin{table}[ht!] \centering 
  \caption{Sample descriptive statistics for 212 matches in German Bundesliga seasons 2017/18 and 2018/19 featuring a 1-1 scoreline} 
  \label{tab:descriptives} 
\begin{tabular}{@{\extracolsep{5pt}}lcccc} 
\\[-1.8ex]\hline 
 \\[-1.8ex] 
  & \multicolumn{1}{c}{Mean} & \multicolumn{1}{c}{St.\ dev.} & \multicolumn{1}{c}{Min.\ } & \multicolumn{1}{c}{Max.\ } \\ 
\hline \\[-1.8ex] 
\textit{relstake} (equalising team) & 0.470 & 0.253 & 0.049 & 0.960 \\ 
\textit{relstake} (conceding team) & 0.372 & 0.247 & 0.012 & 0.911 \\ 
\textit{impprob} (equalising team) & 0.357 & 0.188 & 0.037 & 0.909 \\ 
\textit{impprob} (conceding team) & 0.445 & 0.192 & 0.050 & 0.926 \\ 
\textit{minute} & 46.87 & 20.30 & 5 & 84 \\ 
\textit{redcard} & 0 & 0.195 & $-$1 & 1 \\ 
\hline 
\end{tabular} 
 \caption*{\footnotesize Notes.- \textit{relstake}: the relative betting stakes placed in the minute after the equaliser for the 1-1 scoreline; \textit{impprob}: the inverse of decimal odds at kick-off; \textit{minute}: the minute of the match when the 1-1 equaliser arrived; and \textit{redcard}: the difference in red cards received between the equalising and conceding teams before the 1-1 equaliser.}
\end{table}

\FloatBarrier
\section{The Impact of Momentum on Match Outcomes and Betting Markets}

\noindent While the summary statistics in the previous section suggest that bettors tend to adjust their bets in accordance with the match dynamics, an in-depth analysis is needed to understand whether momentum predicts (i) the final match outcome, (ii) how the bookmaker sets prices, and (iii) how bettors respond. Our approach is in three parts. First, we analyse whether the sequence of goals (momentum) in getting to a 1-1 scoreline impacts the eventual overall outcome of matches. If it does, then betting on that momentum could be a rational action if the bookmaker does not reflect it in their pricing. Second, we determine whether the bookmaker adjusts its pricing in line with what might be perceived as momentum. Third and finally, we study the stakes placed after the 1-1 equaliser to determine whether betting activity is affected by what might either be or only appear to be momentum.

\subsection{Do Equalising Goals Generate Momentum?}
\noindent In the absence of any momentum and everything else equal (i.e., the same balance of winning probabilities and thus team strengths as at kick-off, the minute in the match, and the number of red cards received), the probability of a team winning a football match would be identical after conceding the equalising goal at 1-1 compared with scoring it. On the contrary, a greater probability to go on and subsequently win the match when equalising instead of conceding would indicate a genuine sense of momentum.

We modify our dataset for our empirical approach. Each match appears twice in the estimation samples, once from the perspective of the equalising team and once from the perspective of the conceding team. The response variable in this analysis is $\textit{win}_{i,m}$, where $\textit{win}_{i,m}=1$ if the considered team $i$ actually won match $m$ and is zero otherwise. The main explanatory variable of interest is $\textit{equaliser}_{i,m}$, where $\textit{equaliser}_{i,m}=1$ when the considered team $i$ scored the equalising goal in match $m$ and $\textit{equaliser}_{i,m}=0$ if they conceded. For the control variables, $\textit{impprob}_{i,m}$ covers the winning chances of team $i$ prior to match $m$. The higher is $\textit{impprob}_{i,m}$, the higher we would expect the winning chances to be right after the equalising goal. $\textit{minute}_m$ captures the minute the equalising goal was scored. The later in the match that the equaliser is scored, the more likely the match ends in a draw and the less likely either team is going to win. $\textit{redcard}_{i,m}$ is the difference in the number of red cards between the teams. If $\textit{redcard}_{i,m}>0$, then the team received more red cards than the opponent and hence has less players on the field. We expect the winning chances of a team to decrease as $\textit{redcard}_{i,m}$ increases. As such, we model whether a team won the match using logistic regression:
\begin{align}
    \text{logit}\big(\Pr(win_{i,m} = 1) \big) =& \beta_0 + \beta_1 \cdot impprob_{i,m} + \beta_2 \cdot equaliser_{i,m} + \notag \\ & \beta_3 \cdot minute_{m} + \beta_4 \cdot redcard_{i,m} \ . \label{eq:one}
\end{align}
Since each match appears twice in the estimations, we cluster standard errors at the match level. 

Table~\ref{tab:momentum} displays the results for the full sample of 212 matches (Column I) and for observations of the first (Column II) and second half (Column III), respectively. We do not find evidence for equalising goals generating momentum in the full sample. All the control variables generate coefficients with expected signs for the likelihood of a team going on to win after a 1-1 equaliser: positive for the pre-match expectations of a win according to odds; negative for a late equaliser; and negative for having received more red cards than the opponent. Checking whether momentum exists in the first or second half separately, we again do not find any evidence that equalising goals generate momentum in any of these cases.

In Appendix~Table~\ref{tab:momentum_robustness}, we show that the main results in column (I) of Table~\ref{tab:momentum} are robust to extensions of the model given by Equation~\eqref{eq:one}, including: a squared term for the minute of the equalising goal; interacting the minute of the equaliser and $redcard_{i,m}$, which is insignificant; interacting $impprob_{i,m}$ and $equaliser_{i,m}$, in case there is momentum only for either equalising favourites or longshots, which there is not; and interacting $equaliser_{i,m}$ and $minute_{m}$, in case there is momentum only for either late or early goals, which there is not.

\begin{table}[ht!] 
  \caption{Does scoring momentum impact match outcomes?}
   \centering
  \label{tab:momentum} 
\begin{tabular}{@{\extracolsep{5pt}}lccc} 
\\[-1.8ex]\hline 
\\[-1.8ex] 
  & \multicolumn{3}{c}{\textit{Timing of equaliser for 1-1}} \\ 
\cline{2-4} 
\\[-1.8ex] & Any time & First half & Second half \\ 
\\[-1.8ex] & (I) & (II) & (III) \\ 
\hline \\[-1.8ex] 
 \textit{impprob} ($\beta_1$) & 0.045$^{***}$ & 0.035$^{***}$ & 0.058$^{***}$ \\ 
  & (0.008) & (0.010) & (0.013) \\ 
  & & & \\ 
 \textit{equaliser} ($\beta_2$) & 0.115 & $-$0.010 & 0.257 \\ 
  & (0.286) & (0.397) & (0.415) \\ 
& & & \\ 
 \textit{minute} ($\beta_3$) & $-$0.014$^{***}$ & $-$0.019$^{*}$ & $-$0.030$^{**}$ \\ 
  & (0.005) & (0.010) & (0.014) \\ 
  & & & \\ 
 \textit{redcard} ($\beta_4$) & $-$2.268$^{***}$ & (no red cards)  & $-$2.411$^{***}$ \\ 
  & (0.646) &  & (0.672) \\ 
  & & & \\ 
 Constant ($\beta_0$) & $-$2.101$^{***}$ & $-$1.564$^{***}$ & $-$1.672 \\ 
  & (0.446) & (0.562) & (1.096) \\ 
  & & & \\ 
\hline 
$N$ of matches & 212 & 101 & 111 \\
$N$ of observations & 424 & 202 & 222  \\
McFadden $R^2$ & 0.132 & 0.084 & 0.191 \\ 
\hline
\end{tabular} 
\caption*{\footnotesize Notes.- Logistic regression estimates of Equation~\eqref{eq:one}. \newline $^{***}$, $^{**}$, $^{*}$ indicate significance from zero of the model coefficients at the 1\%, 5\% and 10\% levels, respectively, two-sided tests, with standard errors in parentheses that account for match-level clustering.}
\end{table} 
\FloatBarrier
\subsection{Does the Bookmaker `Believe' in Momentum?}
\noindent While we find no evidence in our sample of matches that momentum affects the likelihood of match outcomes, the bookmaker could still systematically alter odds according to the sequence of goals in a 1-1 scoreline for two reasons. First though unlikely, the price setting could be biased because the bookmaker believes in the impact of momentum on the final match outcomes. Second, it could anticipate that bettors believe in momentum and adjust betting odds accordingly to secure profits.

To check whether bookmaker pricing of the win outcomes is affected by the sequence of goals scored up to an equaliser, we consider the bookmaker's posted \textit{odds} for each team $i$ to win in the minute after the 1-1 scoreline in match $m$. We use the same control variables as for the match outcome in Equation~\eqref{eq:one} and consider the following linear regression model: 
\begin{align}
    odds_{i,m}  = & \,  \beta_0 + \beta_1 \cdot impprob_{i,m} +  \beta_2 \cdot equaliser_{i,m} +  \beta_3 \cdot minute_m \notag\\ & +  \beta_4 \cdot redcard_{i,m} + u_{i,m} \ . \label{eq:two}
\end{align}

Table~\ref{tab:bookmaker} reports the estimated parameters, again for both the full sample and sub-samples for first and second half equalisers. For all model formulations, the identity of who scored the $\textit{equaliser}$ has an insignificant effect on the bookmaker win odds. The estimated direction of the control variable effects on the post-equaliser betting odds are all in line with our expectations. The pre-match expectations of a win for a team, proxied by $\textit{impprob}_{i,m}$, can significantly and positively explain their odds to win after the equaliser. The later in the match that the equaliser was scored, the lower the team's subsequent odds to secure a win, as the draw is the most probable final outcome. The difference in the number of red cards received prior to the equaliser reduced the winning odds of a team.

In columns (I)-(IV) of Appendix~Table~\ref{tab:bookmaker_robustness}, we show that the main results in column (I) of Table~\ref{tab:bookmaker} are robust to extensions of the model given by Equation~\eqref{eq:two}, including: a squared term for the minute of the equalising goal; interacting the minute of the equaliser and $redcard_{i,m}$, which is insignificant; interacting $impprob_{i,m}$ and $equaliser_{i,m}$, in case there is evidence that the bookmaker only alters odds according to whether it was either the pre-match favourite or longshot that scored the equaliser, which there is not; and interacting $equaliser_{i,m}$ and $minute_{m}$, in case there is evidence that the bookmaker only alters odds according to whether it was a late or early equalising goal, which there is not. In columns (V)-(VIII) of Appendix~Table~\ref{tab:bookmaker_robustness}, comparable results are shown that use the odds-implied probability of a win after the equaliser as the response variable, instead of the decimal odds, thus showing that our results are not sensitive to this choice.
\begin{table}[ht!] \centering 
  \caption{Do bookmaker odds for the win reflect momentum?
  } 
  \label{tab:bookmaker} 
  \scalebox{0.9}{
\begin{tabular}{@{\extracolsep{5pt}}lccc} 
\\[-1.8ex]\hline 
 \\[-1.8ex] 
  & \multicolumn{3}{c}{\textit{Timing of equaliser for 1-1}} \\ 
\cline{2-4} 
\\[-1.8ex] & Any time & First half & Second half \\ \\[-1.8ex] & (I) & (II) & (III)  \\
\hline \\[-1.8ex] 
 \textit{impprob} ($\beta_1$) & $-$0.100$^{***}$ & $-$0.109$^{***}$ & $-$0.090$^{***}$ \\ 
  & (0.004) & (0.007) & (0.004) \\ 
  & & & \\ 
 \textit{equaliser} ($\beta_2$) & $-$0.017 & $-$0.255 & 0.204$^{*}$ \\ 
  & (0.128) & (0.223) & (0.122) \\ 
  & & & \\ 
 \textit{minute} ($\beta_3$) & 0.011$^{**}$ & $-$0.008 & 0.040$^{***}$ \\ 
  & (0.005) & (0.012) & (0.009) \\ 
  & & & \\ 
 \textit{redcard} ($\beta_4$) & 1.931$^{***}$ & (no red cards)  & 1.902$^{***}$ \\ 
  & (0.300) &  & (0.317) \\ 
  & & & \\ 
 Constant ($\beta_0$) & 7.118$^{***}$ & 8.158$^{***}$ & 4.820$^{***}$ \\ 
  & (0.438) & (0.596) & (0.594) \\ 
  & & & \\ 
\hline \\[-1.8ex] 
$N$ of matches & 212 & 101 & 111 \\
$N$ of observations & 424 & 202 & 222  \\
$R^{2}$ & 0.563 & 0.506 & 0.709 \\ 
\hline 
    \end{tabular}}
\caption*{\footnotesize Notes.- Estimates of Equation~\eqref{eq:two}. $^{***}$, $^{**}$, $^{*}$ indicate significance from zero of the model coefficients at the 1\%, 5\% and 10\% levels, respectively, two-sided tests, with standard errors in parentheses that account for match-level clustering.}
\end{table}
\FloatBarrier
\subsection{Do Bettors `Believe' in Momentum?}

\noindent The results from the previous two parts of our analysis provide evidence for neither teams generally gaining momentum after scoring a 1-1 equaliser nor the bookmaker pricing the win according to which team scored last. In the third part of our analysis, we study the relative stakes placed by bettors after 1-1 equalisers, to investigate whether or not bettors believe there is value in momentum and bet more money than they should on the equalising teams.

The descriptives in Table~\ref{tab:descriptives} above already indicate that the relative stakes placed on the equalising team are substantially higher compared to the team conceding the 1-1 equaliser. To extend this exploratory analysis, we split our sample according to money bet on strong favourites (pre-match odds $<$ 1.5), moderate favourites (pre-match odds between 1.5 and 2.7), moderate longshots (pre-match odds between 2.7 and 4) and strong longshots (pre-match odds~$>$~4). As our previous results did not indicate any significant momentum effects for the match outcome nor corresponding pricing response by the bookmaker, the relative stakes for both equalising and conceding teams should be very similar if bettors behaved rationally. However, for the four subsamples considered, the mean relative stakes placed on the team that scored/conceded the equaliser are as follows:

\begin{itemize}
	\item[-] strong favourites: 0.779 (scored), 0.591 (conceded)
	\item[-] moderate favourites: 0.542 (scored), 0.415 (conceded)
	\item[-] moderate longshots: 0.431 (scored), 0.248 (conceded)
	\item[-] strong longshots: 0.290 (scored), 0.157 (conceded)
\end{itemize}

These summary statistics indicate substantial differences in betting activity between stakes placed on the equalising and conceding teams. In addition to these means, Appendix~Figure~\ref{fig:hists} shows the corresponding histograms of relative stakes for the four subsamples. For bets on strong pre-match favourites, relative stakes are generally larger if they score rather than concede the 1-1 equaliser. In this case, there is some support for bettors acting sensibly to the return on investment from following a hypothetical betting strategy of always betting on a clear pre-match favourite that just scored the equaliser, although from a small sample. Such a strategy would have yielded a return on investment (ROI) of 0.6\% (20 bets, 13 won), compared with the reverse strategy of always betting the same amount on clear pre-match favourites that just conceded the equaliser, which yields an ROI of -13.9\% (29 bets, 17 won). For context, the average overround (sometimes called the `vig' or `profit margin') implied by the bookmaker odds in our data after the equalising goal is 7.9\%.

Similar patterns can be observed for stakes placed on pre-match moderate favourites, moderate longshots, and strong longshots. Reinforcing the patterns observed in Figure~\ref{fig:example_data} for the example match earlier, there is clear exploratory evidence that the relative stakes for the team that scores at 1-1 are, on average, substantially larger in the aftermath, thus suggesting that bettors believe betting on momentum, conditional on the prices offered, will be profitable for them. They were clearly wrong, at least if they were following the simple strategy of always betting the same amount on the equalising team. The ROIs from always backing equalising pre-match moderate favourites, moderate longshots, and strong longshots in such a way, in our sample, would have been -20.1\% (70 bets, 23 won), -7.4\% (50 bets, 12 won), and -23.3\% (72 bets, 10 won), respectively.

To support what appears to be convincing exploratory findings, i.e., that betting activity tends to follow the team with apparent momentum, we use an econometric approach. To explain $\textit{relstake}_{i,m}$ -- the share of volume bet on a particular team $i$ to win in the minute after the equalising goal in match $m$ -- we estimate the following model:
\begin{align}
   relstake_{i,m} = & \beta_0 + \beta_1 \cdot startodds_{i,m} + \beta_2 \cdot equaliser_{i,m} + \beta_3 \cdot minute_m \notag\\ & + 
    \beta_4 \cdot prerelstake_{i,m} + \beta_5 \cdot redcard_{i,m} + u_{i,m} \ , \label{eq:three}
\end{align}
where control variables once again include $minute_m$ and $redcard_{i,m}$. We also control for the starting odds of a win prior to the match for team $i$, $startodds_{i,m}$, to model whether bettors tend to bet on pre-match favourites after an equaliser. Further, we include the relative stakes in the one minute prior to the equaliser, $prerelstake_{i,m}$ to model the general tendency of the market to favour betting on one team over the other possible match outcomes. 

The estimation results in Table~\ref{tab:bettors} show that bettors generally tend to back pre-match favourites over longshots after an equaliser, though only significantly so in the second half after conditioning on the relative stakes prior to the first goal of the match. Again, the later the equaliser is scored, the lower are the chances that one team goes on to win, and hence the lower the relative stakes on a win as the final outcome. As red cards reduce the chances of winning, bettors tend to back wins less for undermanned teams (unless there are 15 or fewer minutes remaining in the match). The relative stakes placed on a team prior to an equalising goal significantly predict the relative stakes after, though the coefficient for this variable in the model is also significantly less than one. Most importantly, column (I) of Table~\ref{tab:bettors} supports the exploratory findings from above and shows that significantly and substantially higher betting volumes tend to back teams that score the equaliser compared to teams that concede. The model estimates show that the relative stakes placed on a team to win are on average 12.7 percentage points higher in the minute after they scored the equaliser compared with if they had instead conceded, remembering also that absolute betting volumes tend to double in our sample after the equaliser compared with the three minutes before. 
For the model shown in column (I), fixing all control variables at their respective means, the stakes placed on the equalising team are 35.7\% higher than those placed on the conceding team. The corresponding increase for the model shown in column (II) is substantially higher at 46.5\%. The goodness of fit of our model is also promising, with an $R^2$ of more than 0.6, and our findings hold for sub-samples according to when the equaliser went in, shown in columns (II-III) of Table~\ref{tab:bettors}. In particular, an increase in the relative stakes placed on the equalising team is especially pronounced in the second half of matches, at which times relative stakes on teams that have just equalised to win are 19.4 percentage points greater than on the conceding teams, holding everything else in the models equal.

\begin{table}[ht!] \centering 
  \caption{Do bettors follow the apparent momentum?} 
  \label{tab:bettors} 
  \scalebox{0.9}{
\begin{tabular}{@{\extracolsep{5pt}}lcccc} 
\\[-1.8ex]\hline 
 \\[-1.8ex] 
   & \multicolumn{4}{c}{\textit{Timing of equaliser for 1-1}} \\ 
\cline{2-5}
\\[-1.8ex] & Any time & Any time & First half & Second half  \\ \\[-1.8ex] & (I) & (II) & (III) & (IV) \\ 
\hline \\[-1.8ex] 
\textit{startodds} ($\beta_1$) & $-$0.037$^{***}$ & $-$0.004 & 0.002 & $-$0.013$^{**}$ \\ 
  & (0.008) & (0.004) & (0.004) & (0.006) \\ 
  & & & & \\ 
 \textit{equaliser} ($\beta_2$)  & 0.127$^{***}$ & 0.159$^{***}$ & 0.123$^{***}$ & 0.194$^{***}$ \\ 
  & (0.028) & (0.020) & (0.029) & (0.027) \\ 
  & & & & \\ 
 \textit{minute} ($\beta_3$) & $-$0.002$^{***}$ & $-$0.001$^{***}$ & $-$0.0003 & $-$0.003$^{***}$ \\ 
  & (0.000) & (0.000) & (0.0004) & (0.001) \\ 
  & & & & \\ 
 \textit{prerelstake} ($\beta_4$)  &  & 0.725$^{***}$ & 0.777$^{***}$ & 0.653$^{***}$ \\ 
  &  & (0.040) & (0.052) & (0.061) \\ 
  & & & & \\ 
 \textit{redcard} ($\beta_5$) & $-$0.169$^{***}$ & $-$0.189$^{***}$ & (no red cards) & $-$0.190$^{***}$ \\ 
  & (0.052) & (0.069) &  & (0.067) \\ 
  & & & & \\ 
 Constant ($\beta_0$) & 0.562$^{***}$ & 0.087$^{***}$ & 0.031 & 0.262$^{***}$ \\ 
  & (0.032) & (0.033) & (0.038) & (0.064) \\ 
  & & & & \\
\hline \\[-1.8ex] 
$N$ of matches & 211 & 211 & 100 & 111 \\ 
$N$ of observations & 422 & 422 & 200 & 222 \\
$R^{2}$ & 0.229 & 0.621 & 0.644 & 0.623 \\
\hline 
\end{tabular}}
\caption*{\footnotesize Notes.- Estimates of Equation~\eqref{eq:three}. $^{***}$, $^{**}$, $^{*}$ indicate significance from zero of the model coefficients at the 1\%, 5\% and 10\% levels, respectively, two-sided tests, with standard errors in parentheses that account for match-level clustering. The sample contains 211 instead of 212 matches since in one match no stakes were placed in the next minute after the 1-1 equaliser.}
\end{table} 

In Appendix~Table~\ref{tab:bettors_robustness}, we show that the main results in column (I) of Table~\ref{tab:bettors} are robust to extensions of the model given by Equation~\eqref{eq:three}, including: squared terms for the minute of the equalising goal and the relative stakes in the one minutes prior to the equaliser; interacting the minute of the equaliser and $redcard_{i,m}$, which is insignificant; interacting $startodds_{i,m}$ and $equaliser_{i,m}$, in case there is evidence that bettors tend to back pre-match equalising favourites more greatly than longshots, which there is not; and interacting $equaliser_{i,m}$ and $minute_{m}$. Appendix~Table~\ref{tab:bettors_robustness} further includes a robustness check with relative stakes three minutes after the equaliser as the response variable (shown in the first column). In that model formulation, stakes on the win are also substantially larger for teams that score the equaliser than for teams that concede.

\section{Conclusion}
\noindent This paper tackles the question of how gamblers adjust their risk-taking behaviour to the possibility of perceived momentum in the value of a state-contingent asset, viewed through observing betting activity within football matches. We use a novel and rich dataset from a large and well-known international bookmaker, focusing on betting markets just after 1-1 equalisers are scored during matches in the German Bundesliga. We analyse whether the sequence of scoring impacts the final match outcome, the price setting by the bookmaker, and ultimately the amount and direction of betting activity. On the sequence of scoring, we hypothesise that the equalising team has gained momentum. However, our results suggest that on average the sequence of the goals leading to the 1-1 does not influence the ultimate winning chances of a team or the odds setting by the bookmaker. Still, there is convincing evidence that bettors believe in the value of momentum, as considerably higher stakes are placed on the teams that have just equalised to eventually win, compared with the teams that conceded. Such perceived value in momentum among bettors does not translate into profits, as always betting on the team with momentum on average leads to significant negative returns.

To the best of our knowledge, this paper is the first that can cleanly isolate a singular event (in our case the equaliser in a football match) that creates momentum (for the team that scored the goal) and influences investor behaviour. It demonstrates that while the order of goals is irrelevant to the game outcome, bettors heavily believe in its importance. This indicates that investors on betting markets indeed have difficulty quantifying the relevance of even the most important game situation in a football match. Still, there are some potential limitations to our approach. First, since the data are accumulated over all bettors, it is impossible to connect multiple bets by a single bettor over time. Potentially, some bettors stake money on one particular outcome during the match before later staking on a different outcome, to hedge the first bet. Therefore, betting after the equaliser might in part reflect activity by the very same bettors prior to the goal. The impacts of such strategic and dynamic betting behaviour on our findings should be quite low though, as it would most likely only make sense when a bettor had staked money after but not prior to the first goal. Second, while our analysis covers rich betting data, the detail level of in-play statistics that we could reliably link to them is low. While goals (and red cards) are by far the most meaningful events to impact football match outcomes, others, such as yellow cards, substitutions, shots on target or corner kicks, could impact how bettors perceive the momentum of teams. Last, our match-by-match analysis neglects any cross-match momentum or how teams handle certain in-play situations they have experienced in the past. If one team scored the equaliser in a previous match to later go on and win, then bettors could predict the re-occurrence of such a dynamic in the next match and bet accordingly right after an equaliser.

The above mentioned limitations could and should be tackled in future works, as knowledge of the drivers of betting behaviour is still in its infancy --- almost all the literature is focused only on prices (odds). We believe that betting markets are valuable settings to understand price setting and behavioural biases that affect risk-taking. The availability of both odds and actual staked amounts enables investigation of how bettors respond to prices and make potentially biased investment decisions.

\clearpage

\clearpage


\newpage

\titleformat{\section}{\large\bfseries}{\appendixname~\thesection .}{0.5em}{}

\begin{appendices}
\section{Additional Tables -- Model Robustness Checks}

\renewcommand\thetable{\thesection\arabic{table}}
\setcounter{table}{0}

\begin{table}[ht!] 
  \caption{Robustness checks on ``Does scoring momentum impact match outcomes?''}
   \centering
  \label{tab:momentum_robustness} 
\begin{tabular}{@{\extracolsep{5pt}}lcccc} 
\\[-1.8ex]\hline 
\\[-1.8ex] 
  & \multicolumn{4}{c}{\textit{Timing of equaliser for 1-1}} \\ 
\cline{2-5} 
\\[-1.8ex] & Any time & Any time & Any time & Any time \\ 
\\[-1.8ex] & (I) & (II) & (III) & (IV) \\ 
\hline \\[-1.8ex] 
 \textit{impprob} & 0.045$^{***}$ & 0.045$^{***}$ & 0.046$^{***}$ & 0.047$^{***}$ \\ 
  & (0.008) & (0.008) & (0.010) & (0.010) \\ 
  & & & & \\ 
 \textit{equaliser} & 0.115 & 0.108 & 0.260 & 0.407 \\ 
  & (0.286) & (0.287) & (0.557) & (1.340) \\ 
  & & & & \\ 
 \textit{minute} & 0.004 & 0.001 & 0.001 & 0.0004 \\ 
  & (0.020) & (0.019) & (0.019) & (0.035) \\ 
  & & & & \\ 
 $\textit{minute}^2$ & $-$0.0002$^{***}$ & $-$0.0002$^{***}$ & $-$0.0002$^{***}$ & $-$0.0001$^{***}$ \\ 
  & (0.000) & (0.000) & (0.000) & (0.000) \\ 
  & & & & \\ 
 \textit{redcard} & $-$2.304$^{***}$ & 30.407 & 30.159 & 33.044 \\ 
  & (0.645) & (35.605) & (35.440) & (36.653) \\ 
  & & & & \\ 
 \textit{minute} $\cdot$ \textit{redcard} &  & $-$1.048 & $-$1.041 & $-$1.131 \\ 
  &  & (1.080) & (1.075) & (1.115) \\ 
  & & & & \\ 
 $\textit{minute}^2$ $\cdot$ \textit{redcard} &  & 0.008 & 0.008 & 0.009 \\ 
  &  & (0.008) & (0.008) & (0.008) \\ 
  & & & & \\ 
 \textit{impprob} $\cdot$ \textit{equaliser}  &  &  & $-$0.004 & $-$0.003 \\ 
  &  &  & (0.011) & (0.011) \\ 
  & & & & \\ 
 \textit{minute} $\cdot$ \textit{equaliser} &  &  &  & 0.002 \\ 
  &  &  &  & (0.059) \\ 
  & & & & \\ 
 $\textit{minute}^2$ $\cdot$ \textit{equaliser} &  &  &  & $-$0.0001 \\ 
  &  &  &  & (0.001) \\ 
  & & & & \\ 
 Constant & $-$2.417$^{***}$ & $-$2.364$^{***}$ & $-$2.451$^{***}$ & $-$2.531$^{***}$ \\ 
  & (0.541) & (0.535) & (0.605) & (0.842) \\ 
  & & & & \\ 
\hline 
$N$ of matches & 212 & 212 & 212 & 212  \\
$N$ of observations & 424 & 424 & 424 & 424 \\
\hline
\end{tabular} 
\caption*{\footnotesize Notes.- Logistic regression estimates of Equation~\eqref{eq:one} with added control variables and interactions. See Table~\ref{tab:momentum}. \newline $^{***}$, $^{**}$, $^{*}$ indicate significance from zero of the model coefficients at the 1\%, 5\% and 10\% levels, respectively, two-sided tests, with standard errors in parentheses that account for match-level clustering.}
\end{table}

\begin{table}[!htbp] \centering 
  \caption{Robustness checks on ``Do bookmaker odds for the win reflect momentum?''
  } 
  \label{tab:bookmaker_robustness} 
  \scalebox{0.7}{
\begin{tabular}{@{\extracolsep{5pt}}lcccccccc}
\\[-1.8ex]\hline 
 \\[-1.8ex] 
  & \multicolumn{7}{c}{\textit{Timing of equaliser for 1-1}} \\ 
\cline{2-9} 
& \multicolumn{4}{c}{\textit{response: \textit{odds} after 1-1}} & \multicolumn{4}{c}{\textit{response: \textit{impprob} after 1-1}}  \\ 
\cline{2-5}  \cline{6-9}  
\\[-1.8ex] & Any time & Any time & Any time & Any time & Any time & Any time & Any time & Any time\\ 

\\[-1.8ex] & (I) & (II) & (III) & (IV) & (V) & (VI) & (VII) & (VIII) \\ 
\hline \\[-1.8ex] 
 \textit{impprob} & $-$0.100$^{***}$ & $-$0.100$^{***}$ & $-$0.099$^{***}$ & $-$0.099$^{***}$ & 0.008$^{***}$ & 0.008$^{***}$ & 0.008$^{***}$ & 0.008$^{***}$ \\ 
  & (0.004) & (0.004) & (0.010) & (0.010) & (0.0002) & (0.0002) & (0.0002) & (0.0002) \\ 
  & & & & & & & & \\ 
 \textit{equaliser} & $-$0.017 & $-$0.031 & 0.110 & $-$1.052 & $-$0.004 & $-$0.003 & $-$0.005 & 0.037 \\ 
  & (0.129) & (0.128) & (0.758) & (0.959) & (0.008) & (0.008) & (0.008) & (0.025) \\ 
  & & & & & & & & \\ 
 \textit{minute} & $-$0.041$^{**}$ & $-$0.041$^{**}$ & $-$0.041$^{*}$ & $-$0.062$^{**}$ & 0.002$^{***}$ & 0.002$^{***}$ & 0.002$^{***}$ & 0.003$^{***}$ \\ 
  & (0.021) & (0.021) & (0.021) & (0.026) & (0.0002) & (0.0002) & (0.0002) & (0.001) \\ 
  & & & & & & & & \\ 
 $\textit{minute}^2$ & 0.001$^{***}$ & 0.001$^{***}$ & 0.001$^{***}$ & 0.001$^{***}$ & $-$0.00004$^{***}$ & $-$0.00004$^{***}$ & $-$0.00004$^{***}$ & $-$0.0001$^{***}$ \\ 
  & (0.0002) & (0.0002) & (0.0002) & (0.0003) & (0.00000) & (0.00000) & (0.00000) & (0.00001) \\ 
  & & & & & & & & \\ 
 \textit{redcard} & 1.931$^{***}$ & 18.310$^{*}$ & 18.319$^{*}$ & 17.126 & $-$0.133$^{***}$ & $-$0.989$^{***}$ & $-$0.989$^{***}$ & $-$1.072$^{***}$ \\ 
  & (0.300) & (10.615) & (10.634) & (11.092) & (0.016) & (0.200) & (0.200) & (0.196) \\ 
  & & & & & & & & \\ 
 \textit{minute} $\cdot$ \textit{redcard} &  & $-$0.471 & $-$0.471 & $-$0.437 &  & 0.023$^{***}$ & 0.023$^{***}$ & 0.025$^{***}$ \\ 
  &  & (0.340) & (0.340) & (0.354) &  & (0.006) & (0.006) & (0.006) \\ 
  & & & & & & & & \\ 
 $\textit{minute}^2$ $\cdot$ \textit{redcard} &  & 0.003 & 0.003 & 0.003 &  & $-$0.0001$^{***}$ & $-$0.0001$^{***}$ & $-$0.0002$^{***}$ \\ 
  &  & (0.003) & (0.003) & (0.003) &  & (0.00004) & (0.00004) & (0.00004) \\ 
  & & & & & & & & \\ 
 \textit{impprob} $\cdot$ \textit{equaliser} &  &  & $-$0.004 & $-$0.003 &  &  & 0.0001 & 0.00005 \\ 
  &  &  & (0.017) & (0.017) &  &  & (0.0001) & (0.0001) \\ 
  & & & & & & & & \\ 
  \textit{minute} $\cdot$ \textit{equaliser} &  &  &  & 0.042$^{*}$ &  &  &  & $-$0.002 \\ 
  &  &  &  & (0.023) &  &  &  & (0.001) \\ 
  & & & & & & & & \\ 
 $\textit{minute}^2$ $\cdot$ \textit{equaliser} &  &  &  & $-$0.0003 &  &  &  & 0.00002 \\ 
  &  &  &  & (0.0003) &  &  &  & (0.00002) \\ 
  & & & & & & & & \\ 
 Constant & 8.071$^{***}$ & 8.082$^{***}$ & 8.005$^{***}$ & 8.592$^{***}$ & 0.059$^{***}$ & 0.058$^{***}$ & 0.059$^{***}$ & 0.038$^{**}$ \\ 
  & (0.581) & (0.583) & (0.778) & (0.889) & (0.010) & (0.010) & (0.011) & (0.017) \\ 
  & & & & & & & & \\ 
\hline \\[-1.8ex]
$N$ of matches & 212 & 212 & 212 & 212 & 212 & 212 & 212 & 212 \\ 
$N$ of observations & 424 & 424 & 424 & 424 & 424 & 424 & 424 & 424 \\ 
$R^{2}$ & 0.571 & 0.573 & 0.573 & 0.576 & 0.898 & 0.901 & 0.901 & 0.902 \\ 
\hline 
    \end{tabular}}
\caption*{\footnotesize Notes.- columns (I)-(IV) show estimates of Equation~\eqref{eq:two} with added control variables and interactions - see Table~\ref{tab:bookmaker} - and columns (V)-(VIII) show comparable estimates of Equation~\eqref{eq:two} when the response variables is instead the implied probability of a win for team $i$ in match $m$ in the minute after a 1-1 equaliser, instead of the posted decimal odds. \newline Notes: $^{***}$, $^{**}$, $^{*}$ indicate significance from zero of the model coefficients at the 1\%, 5\% and 10\% levels, respectively, two-sided tests, with standard errors in parentheses that account for match-level clustering.}
\end{table}

\begin{table}[ht!] \centering 
  \caption{Robustness checks on ``Do bettors follow the apparent momentum?''} 
  \label{tab:bettors_robustness} 
  \scalebox{0.9}{
\begin{tabular}{@{\extracolsep{5pt}}lccccc}
\\[-1.8ex]\hline 
 \\[-1.8ex] 
   & \multicolumn{5}{c}{\textit{Timing of equaliser for 1-1}} \\ 
\cline{2-5} 
\\[-1.8ex] & Any time & Any time & Any time & Any time & Any time \\ 
\\[-1.8ex] & (3 min.\ after equaliser)  &  &  &  &  \\ 
\hline \\[-1.8ex] 
 \textit{startodds} & $-$0.005 & $-$0.004 & $-$0.004 & $-$0.001 & $-$0.002 \\ 
  & (0.003) & (0.004) & (0.004) & (0.003) & (0.003) \\ 
  & & & & \\ 
 \textit{equaliser} & 0.179$^{***}$ & 0.159$^{***}$ & 0.158$^{***}$ & 0.174$^{***}$ & $-$0.016 \\ 
  & (0.016) &  (0.020) & (0.020) & (0.026) & (0.096) \\ 
  & & & & \\ 
 \textit{minute} & $-$0.001$^{***}$ & 0.002$^{*}$ & 0.002$^{**}$ & 0.002$^{**}$ & $-$0.001 \\ 
  &  (0.000) & (0.001) & (0.001) & (0.001) & (0.002) \\ 
  & & & & \\ 
 $\textit{minute}^2$ & & $-$0.00003$^{***}$ & $-$0.00003$^{***}$ & $-$0.00003$^{***}$ & $-$0.00001$^{***}$ \\ 
  & & (0.000) & (0.000) & (0.000) & (0.000) \\ 
  & & & & \\ 
 \textit{redcard} & $-$0.146$^{***}$ & $-$0.189$^{***}$ & 0.417 & 0.395 & 0.120 \\ 
  & (0.051) & (0.069) & (2.263) & (2.271) & (2.388) \\ 
  & & & & \\ 
 \textit{prerelstake} & 0.732$^{***}$ & 0.724$^{***}$ & 0.602$^{***}$ & 0.613$^{***}$ & 0.618$^{***}$ \\ 
  & (0.034) & (0.040) & (0.085) & (0.084) & (0.085) \\ 
  & & & & \\ 
 $\textit{prerelstake}^2$ & &  & 0.137$^{*}$ & 0.127$^{*}$ & 0.122$^{*}$ \\ 
  & & & (0.070) & (0.071) & (0.071) \\ 
  & & & & \\ 
 \textit{minute} $\cdot$ \textit{redcard} & &  & $-$0.013 & $-$0.013 & $-$0.004 \\ 
  & & & (0.071) & (0.071) & (0.074) \\ 
  & & & & \\ 
 $\textit{minute}^2$ $\cdot$ \textit{redcard} & &  & 0.0001 & 0.0001 & $-$0.00001 \\ 
  & & & (0.001) & (0.001) & (0.001) \\ 
  & & & & \\ 
 \textit{startodds} $\cdot$ \textit{equaliser} & &  &  & $-$0.005 & $-$0.004 \\ 
  & & &  & (0.005) & (0.005) \\ 
  & & & & \\ 
 \textit{minute} $\cdot$ \textit{equaliser} & &  &  &  & 0.006 \\ 
  & & &  &  & (0.005) \\ 
  & & & & \\ 
 $\textit{minute}^2$ $\cdot$ \textit{equaliser} & &  &  &  & $-$0.00004$^{***}$ \\ 
  & & &  &  & (0.000) \\ 
  & & & & \\ 
 Constant & 0.067$^{***}$ & 0.031 & 0.048 & 0.037 & 0.132$^{**}$ \\ 
  & (0.023) & (0.033) & (0.037) & (0.035) & (0.057) \\ 
  & & & & \\ 
\hline \\[-1.8ex] 
$N$ of matches & 196 & 211 & 211 & 211 & 211 \\ 
$N$ of observations & 392 & 422 & 422 & 422 & 422\\ 
$R^{2}$ & 0.731  &  0.624 & 0.628 & 0.628 & 0.638 \\ 
\hline 
\end{tabular}}
\caption*{\footnotesize Notes.- Estimates of Equation~\eqref{eq:three} with added control variables and interactions. See Table~\ref{tab:bettors}. \newline $^{***}$, $^{**}$, $^{*}$ indicate significance from zero of the model coefficients at the 1\%, 5\% and 10\% levels, respectively, two-sided tests, with standard errors in parentheses that account for match-level clustering. The sample contains 211 instead of 212 matches since in one match no stakes were placed in the next minute after the 1-1 equaliser. In the first column, which refers to the stakes placed in the next three minutes after an equaliser, the sample size is slightly smaller as other major events (further goals or red cards) occurred during the next three minutes in 15 matches.}
\end{table} 

\FloatBarrier

\section{Additional Figures --- Relative Stakes Placed on the Apparent Momentum}

\renewcommand\thefigure{\thesection\arabic{figure}}
\setcounter{figure}{0}

\begin{figure}[ht!]
\centering
\caption{Histograms of relative stakes placed on teams to win in the minute following a 1-1 equaliser}
\includegraphics[scale=0.8]{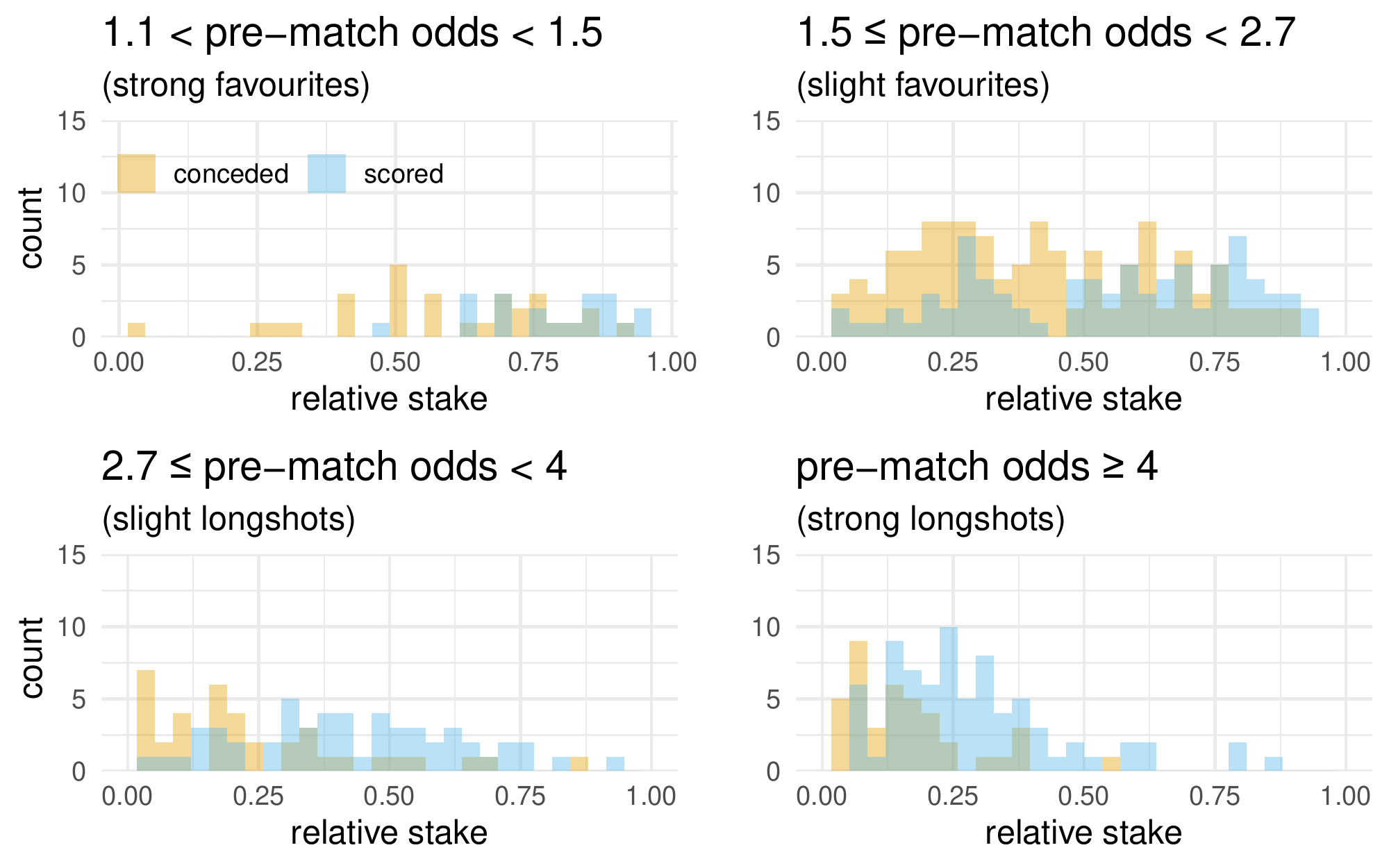}
\caption*{\footnotesize Notes.- Panels show counts of relative stakes on conceding or equalising teams to win over the 212 matches in seasons 2017/18 and 2018/19 that featured a 1-1 scoreline.} 
\label{fig:hists}
\end{figure}

\end{appendices}

\end{document}